\documentclass{article}

\usepackage{arxiv}

\usepackage[utf8]{inputenc} 
\usepackage[T1]{fontenc}    
\usepackage{hyperref}       
\usepackage{amsfonts}       
\usepackage{nicefrac}       
\usepackage{microtype}      
\usepackage{lipsum}

\usepackage{graphicx}
\usepackage{amsmath}
\usepackage{amsthm}
\usepackage{amssymb}
\usepackage{bbm}
\usepackage{bm}
\usepackage{booktabs}
\usepackage{color}
\usepackage{multirow}
\usepackage{natbib}
\usepackage{url}

\newcommand{\R}{\mathbb{R}}

\newcommand{\bx}{\mathbf{x}}
\newcommand{\by}{\mathbf{y}}

\def\btheta{\boldsymbol\theta}

\def\f{{\bf f}}

\title{Scalable modeling of nonstationary covariance functions with non-folding B-spline deformations}

\author{
  Ronaldo~Dias \\
  Department of Statistics\\
  University of Campinas\\
  S\~ao Paulo, Brazil \\
  \texttt{dias@ime.unicamp.br} \\
  \And
  Guilherme~Ludwig \\
  Department of Statistics\\
  University of Campinas\\
  S\~ao Paulo, Brazil \\
  \texttt{gvludwig@ime.unicamp.br} \\
  \And
  Paul~D.~Sampson \\
  Department of Statistics \\
  University of Washington \\
  Washington, USA \\
  \texttt{pds@u.washington.edu} \\
}

\begin{document}







%
%
%
%

\maketitle

\begin{abstract}
We propose a method for nonstationary covariance function modeling, based on the spatial deformation method of \citet{sampson1992nonparametric}, but using a low-rank, scalable deformation function written as a linear combination of the tensor product of B-spline basis.
This approach addresses two important weaknesses in current computational aspects. First, it allows one to constrain estimated 2D deformations to be non-folding (bijective) in 2D. This requirement of the model has, up to now, been addressed only by arbitrary levels of spatial smoothing. Second, basis functions with compact support enable the application to large datasets of spatial monitoring sites of environmental data. An application to rainfall data in southeastern Brazil illustrates the method.
\end{abstract}


\section{Introduction}\label{sec:intro}


Geostatistical methods for spatial and spatio-temporal data are in great demand from fields such as earth and climate sciences, epidemiology and agriculture. A comprehensive overview can be found in \citet{cressie2011statistics}. Spatially stationary processes are commonly used as models in geostatistical applications, but often the assumption of stationarity and isotropic covariance functions are difficult to hold in real applications; see for example, \citet{guttorp1994space}, \citet[pp. 95--101]{le2006statistical}, \citet{damian2001bayesian}. A recent review of methods that allow nonhomogenous covariance models is \citet{schmidt2020flexible}. 

We propose a semiparametric method of nonstationary spatial covariance function that expands upon the work of \citet{sampson1992nonparametric}. Data observed in a set 

\section{Spatial Deformation Model (SDM)}\label{sec:def}

Let $\bx_1, \bx_2, \ldots, \bx_n$ be spatial locations with $\bx_i = (x_{1i}, x_{2i})^t \in G \subset \R^2$ for all $i$ in a geostatistical domain $G,$ and $\by_1, \by_2, \ldots, \by_n,$ with $\by_i = (y_{1i}, y_{2i})^t \in D \subset \R^2$ for all $i.$ Moreover, let $f_\ell : \R^2 \rightarrow \R,$ such that $y_{1i}= f_1(x_{1i},x_{2i}), y_{2i} = f_2(x_{1i}, x_{2i})$ for all $i$ and $f_\ell$ are bijective, differentiable functions, for $\ell = 1,2.$ 

Let $Z$ be a Gaussian random field with spatial covariance function $C.$ For any pair of spatial sites $\bx_i, \bx_{i'},$ we have $Z_{i,t} = Z(\bx_i, t),$ $Z_{i',t} = Z(\bx_{i'}, t),$ with $\mbox{Cov}(Z_{i,t}, Z_{i',t}) = C(\bx_i, \bx_{i'})$ given by
\[
  C(\bx_{i}, \bx_{i'}) = 
\sigma^2 \rho(\|\by_1 - \by_2 \|, \boldsymbol\Delta);
\]

\noindent where $\sigma^2>0,$ and $\rho$ is a stationary, isotropic correlation function with parameters $\boldsymbol\Delta.$ We remark that this model may not be identifiable, since $C$ is invariant for shifts and rigid rotations in $f_1,$ $f_2.$ Moreover, scaling $f_1,f_2$ by a constant $\alpha$ will produce the same $C$ if $\phi$ is also scaled by $\alpha.$ Nevertheless the model can be used for Kriging and spatial interpolation without issues.

\begin{figure}
    \centering
    \includegraphics[scale = .5]{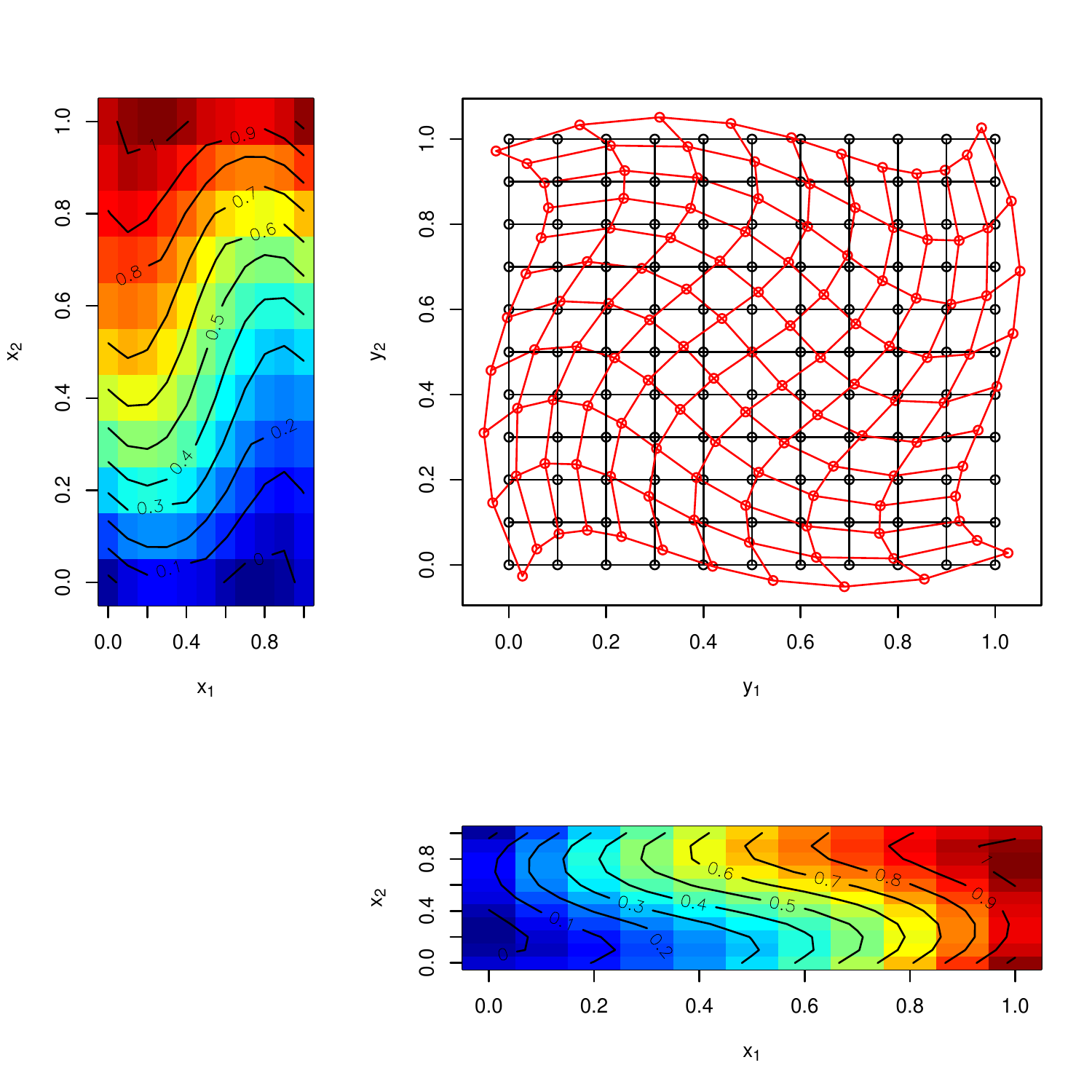}
    \includegraphics[scale = .4]{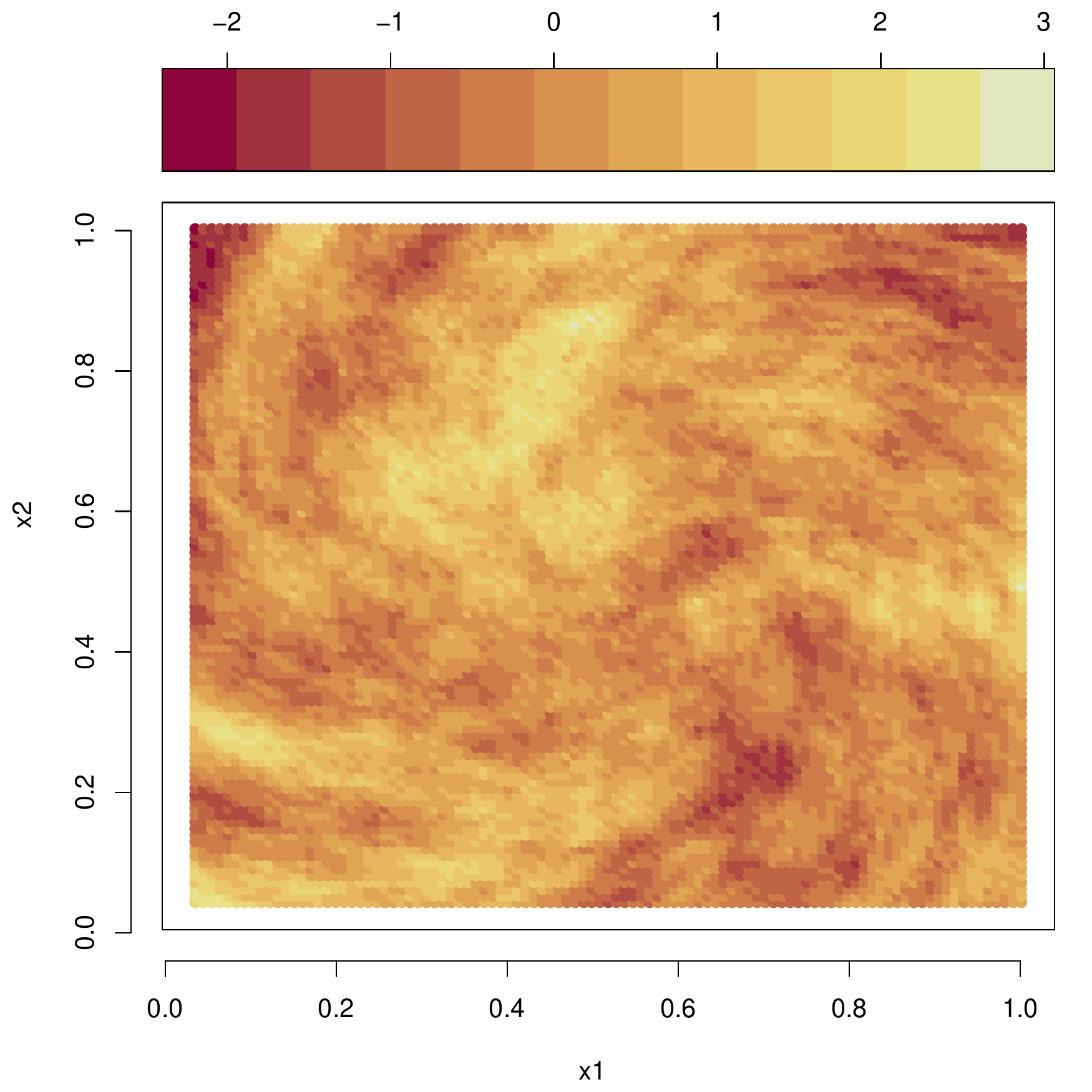}
    \caption{Left panel: regular grid in $\mathcal{G},$ in black, and deformed grid in $\mathcal{D},$ in red. The contour plots in the margins are the the functions $f_1$ and $f_2$ that provide the deformation. Right panel: simulated random field on $\mathcal{G}$ using a deformed exponential covariance function that is stationary on $\mathcal{D}.$}
    \label{fig:example1}
\end{figure}

For example, consider $\mathcal{G} = [0,1]^2$ and $\mathcal{D}$ given by the swirl transformation shown in Figure \ref{fig:example1}. We sampled from a spatio-temporal process on a regular grid of $n=121$ points, with $\mu = 0$ and separable covariance function \[\text{Cov}(Z(\mathbf{x}_1, t_1), Z(\mathbf{x}_2, t_2)) = \exp\{-\|\mathbf{y}_2 - \mathbf{y}_1\|/0.25\}\delta(t_1,t_2),\] where $\mathbf{y}_1 = (f_1(\mathbf{x}_1), f_2(\mathbf{x}_1)),$ $\mathbf{y}_2 = (f_1(\mathbf{x}_2), f_2(\mathbf{x}_2))$ and $\delta(t_1,t_2) = \mathbf{1}\{t_1 = t_2\}.$ A realization of a random field on a fine mesh grid is also shown in Figure \ref{fig:example1}, obtained with the \texttt{RandomFields} package \citep{schlather2015analysis}.

\section{Spatial Deformation Estimation}\label{sec:defest}

Let \[2\gamma(\mathbf{y}_1, \mathbf{y}_2) = \text{Var}(Z(\mathbf{y}_1) - Z(\mathbf{y}_2)),\] then $\gamma$ is the semivariogram of the random field $Z$. In particular, for a stationary, isotropic $Z,$  \[2\gamma(\mathbf{y}, \mathbf{y}+\mathbf{h}) = 2C(0) - 2C(\|\mathbf{h}\|) = g(\|\mathbf{h}\|),\] where $g$ is a positive non-decreasing function of $\|\mathbf{h}\|$. Note valid $g$ functions are conditionally negative definite \citep{cressie1993statistics}.

If there are ways to obtain a sample covariance matrix, such as temporal replicates, then we can construct a sample variogram as \[d_{ij}^2 = 
s_{ii} + s_{jj} - 2s_{ij}.\]

If the variogram is isotropic on a set of artificial coordinates $\mathcal{D}$, then it must be have form \[g(\|\mathbf{y}_i -\mathbf{y}_j\|) = g(h_{ij}),\] where $h_{ij} = \|\mathbf{y}_i-\mathbf{y}_j\|_2.$ Therefore, the variogram entries can be seen as a dispersion metric, with $\hat{g}(h_{ij}) \approx d_{ij}^2.$

\subsection{Non-metric multidimensional scaling}
  
\citet{sampson1992nonparametric} first step is to consider a non-metric multidimensional scaling approach \citep[hereafter, nMDS; see][]{kruskal1964multidimensional,mardia1979multivariate,cox2000multidimensional}. From a set of dispersions $d^2_{ij},$ we seek a monotone transformation $\delta(d_{ij}) = \delta_{ij}$ such that \[\delta(d_{ij}) = \delta_{ij} \approx \|\by_i - \by_j\|\] and \[d^2_{ij} = (\delta^{-1}(\delta_{ij}))^2 \approx g(\|\by_i - \by_j\|),\] and an artificial set of coordinates $\mathbf{y}_i^\ast, \mathbf{y}_j^\ast$ with interpoint Euclidean distances $h^\ast_{ij} = \|\mathbf{y}_i^\ast-\mathbf{y}_j^\ast\|_2$ that minimizes the stress criterion \[\textrm{Stress}(\{h_{ij}^\ast\}_{i>j}) = \sqrt{\frac{\sum_{i>j}\{\delta_{ij}-h^{\ast}_{ij}\}^2}{\sum_{i>j}h_{ij}^{\ast 2}}}.\]

Here we need to impose constraints on $\delta$ such that the function $g$ is conditionally positive definite. \citet{kruskal1964multidimensional} only finds an isotonic transformation, so \citet{sampson1992nonparametric} adapted the optimization algorithm as follows:

\begin{enumerate}

    \item Set k = 1. Find an initial configuration $\by^{\ast (k-1)}_1, \ldots, \by^{\ast (k-1)}_n$ using, for example, Kruskal's isotonic nMDS.

    \item As in Section \ref{sec:tpsmooth} (or Section \ref{sec:cdefest} using our proposed smoother) we smooth the artificial coordinates using the corresponding spline method. Denote them by $\mathbf{f}^{\ast (k-1)}_1, \ldots, \mathbf{f}^{\ast (k-1)}_n.$

    \item Fit a variogram model $g$ to the data using the coordinates $\mathbf{f}^{\ast (k-1)}_1, \ldots, \mathbf{f}^{\ast (k-1)}_n,$ obtaining $\hat{g}_k.$
    
    \item Obtain $\by^{\ast (k)}_2, \ldots, \by^{\ast (k)}_n$ via MDS performed on $[\hat{g}^{-1}_k(d_{ij}^2)]^{1/2}.$
    
    \item Set $k=k+1$ and return to step 2 until convergence.

\end{enumerate}
  
\subsection{Spline smoothing}\label{sec:tpsmooth}

The second step in \citet{sampson1992nonparametric} approach is to find an approximation for $f_1, f_2$ such that $y_{1,i}^* \approx f_1(\mathbf{x}_i),$ $y_{2,i}^* \approx f_2(\mathbf{x}_i).$ This can be done if we approximate the functions $f_1, f_2$ with, for example, thin-plate splines \citep{wahba1980some}. Thin-plate splines are the minimizers of the variational problem \[\min_{f \in \mathcal{W}_2^2} \sum_{i=1}^n (y_i^* -
    f(\mathbf{x}_i))^2 + \lambda \int \left [\left(\frac{\partial^2
          f}{\partial x_1^2}\right)^2 + 2 \left(\frac{\partial^2
          f}{\partial x_1 \partial x_2}\right)^2 +
      \left(\frac{\partial^2 f}{\partial x_2^2}\right)^2 \right]
    \mathrm{d}x_1 \mathrm{d}x_2.\] The solution for a fixed $\lambda$
  is
  \[f_j(x_1, x_2) = \alpha^{(j)}_0 + \alpha^{(j)}_1 x_1 +
    \alpha^{(j)}_2 x_2 + \sum_{i=1}^n \theta_i^{(j)} \varphi_i
    (\|\mathbf{x}_i - \mathbf{x}\|),\] where
  $\varphi(r) = r^2 \log(r)$ and
  $\boldsymbol\alpha, \boldsymbol\theta$ are estimating by plug-in.

\subsection{Tensor product of B-splines}

We propose a different approach for the spline smoothing by making use of low rank approximation (finite approximation)  for the deformation functions. Specifically, let $\bx =(x_1,x_2) \in G \subset \R^2$ and assume we have a collection
of points $(x_{11},x_{21}), \ldots, (x_{1n},x_{2n})$ be a sequence of
points in $G$ . Define a map $\f : G\ \rightarrow D$ 
such that
\begin{equation}
\f (\bx) = \sum_{k=1}^n \psi_k(\bx) \mathbf{y}_k
\end{equation}
where the functions $\psi_k: G \subset \R^2 \rightarrow \R$,  form a partition of unity. 

Particularly, suppose that $\psi_k$ are the B-splines basis functions and there is a integer number $K << n$ such that a map $f : G\ \rightarrow D$
can be represented by 
\begin{equation}
\f (\bx) = \sum_{k=1}^{K} \psi_k(\bx) \btheta_k
\end{equation}

\noindent where $\btheta_k^t = (\theta_k^{(1)}, \theta_k^{(2)})$ and $\f(\mathbf{x})^t = (f_1(\bx),f_2(\bx)).$ Let $\psi_{k}(\mathbf{x}) = B_{k_1}(x_1)B_{k_2}(x_2)$ for some index set $k_1 = 1, \ldots, K_1;$ $k_2 = 1, \ldots, K_2$ such that $K= K_1 \times K_2.$ then the functions $f_\ell$ are well approximated by
\[
\begin{pmatrix}
  f_1 (x_1, x_2) \\
  f_2 (x_1, x_2)
\end{pmatrix} \approx
\begin{pmatrix}
\sum_{{k_1}=1}^{K_1} \sum_{{k_2}=1}^{K_2}
  \theta^{(1)}_{{k_1},{k_2}} B_{k_1}(x_1)B_{k_2}(x_2) \\
  \sum_{{k_1}=1}^{K_1} \sum_{{k_2}=1}^{K_2}
  \theta^{(2)}_{{k_1},{k_2}} B_{k_1}(x_1)B_{k_2}(x_2)
\end{pmatrix}
\]

\noindent where $K_1, K_2$ are fixed positive integers and
$B_{k_1}, B_{k_2}$ are B-spline basis functions \citep[see,
e.g.,][]{ramsay2005functional}. 

To guarantee that the functions $f_\ell$ do not fold onto themselves, we must guarantee that $f_\ell$ is locally invertible and differentiable. In fact, a diffeomorphism is desirable  \citep[see, e.g.,][]{perrin1999modelling}. A necessary condition is that everywhere in the $G$ domain,
\begin{equation}
  |\mathbf{J}| = \frac{\partial f_1}{\partial x_1} \cdot
  \frac{\partial f_2}{\partial x_2} - \frac{\partial f_1}{\partial
    x_2} \cdot \frac{\partial f_2}{\partial x_1} \neq 0,
  \label{eq:constraint}
\end{equation}

\noindent but in practice such constraint is difficult to implement, requiring the evaluation of the Jacobian $\mathbf{J}$ for every pair $(x_1, x_2) \in \mathbb{R}^2.$ We will show in Section \ref{sec:cdefest} that such condition can be translated into a condtion onto the coefficients $\boldsymbol\Theta_\ell = (\theta^{(\ell)}_{{k_1},{k_2}}),$ for $\ell =  1,2.$ Note however that it is enough to ensure $|\mathbf{J}|>0,$ since a change in signs would imply a discontinuous Jacobian.

\subsection{Simultaneous estimation of covariance function and deformation}

The covariance parameters $\boldsymbol\Delta,$ as well as the mean vector $\boldsymbol\mu$ are estimated by maximizing the profile log-likelihood. Without loss of generality set $\boldsymbol\mu = \mathbf{0}.$ Thus our proposed B-spline approach can be estimated as

\begin{enumerate}

\item Given
  $\hat{\boldsymbol\Theta}_1^{(k)}, \hat{\boldsymbol\Theta}_2^{(k)},$
  solve the optimization problem
  \[\hat{\Delta}^{(k)} = \arg\max_\Delta Q_1(\Delta |
    \hat{\boldsymbol\Theta}_1^{(k)},
    \hat{\boldsymbol\Theta}_2^{(k)})\] where
  $Q_1(\Delta|\boldsymbol\Theta_1, \boldsymbol\Theta_2)$ corresponds
  to the optimization target $Q$ seen as a function of $\Delta$ only,
  with parameters $\boldsymbol\Theta_1, \boldsymbol\Theta_2$ fixed.

\item {Given $\hat{\Delta}^{(k)},$ obtain
  $\hat{\boldsymbol\Theta}_1^{(k+1)},
  \hat{\boldsymbol\Theta}_2^{(k+1)}$ such that
  \[
    \begin{aligned} \arg\min_{\boldsymbol\Theta_1,
        \boldsymbol\Theta_2} & \text{tr} \left(\mathbf{R}^t \mathbf{R}
      \right) \\
      \text{s.t.  } & \text{vec}(\boldsymbol\Theta_1)^t
      \mathbf{A}_{i,j}
      \text{vec}(\boldsymbol\Theta_2) > 0, \text{ for } i = 2, \ldots, K, \ j = 2, \ldots, K \\
    \end{aligned}
  \] where $\mathbf{R}_{n \times 2} = \mathbf{Y} - (\mathbf{W}
  \text{vec}(\boldsymbol\Theta_1), \ \mathbf{W}
  \text{vec}(\boldsymbol\Theta_2)),$
  $\mathbf{W}_{n \times K_1K_2} = [\mathbf{b}_2(x_{i,2}) \otimes
  \mathbf{b}_1(x_{i,1})]_{i=1,2,\ldots,n},$ $\mathbf{b}$ are the row
  vector of basis functions and $\otimes$ is the Kronecker product, and for some choice of $\mathbf{A}_{i,j}$ such that $|\mathbf{J}| > 0$ if $\text{vec}(\boldsymbol\Theta_1)^t
      \mathbf{A}_{i,j}
      \text{vec}(\boldsymbol\Theta_2) > 0.$
  
  }

\item Repeat steps 1 and 2 until convergence.
  
\end{enumerate}

We remark that the matrices $\mathbf{W}$ are sparse when using B-splines, therefore solutions to step 2 are scalable. To evaluate the inequality $|\mathbf{J}| > 0,$ consider the following:
write
\[
  f_\ell(x_1, x_2) =
  \begin{pmatrix}
    B_1(x_1) & \cdots & B_{K_1}(x_1)
  \end{pmatrix}
  \begin{pmatrix}
    \theta_{1,1}^{(\ell)} & \cdots & \theta_{1,K_2}^{(\ell)} \\
    \vdots & \ddots & \vdots \\
    \theta_{K_1,1}^{(\ell)} & \cdots & \theta_{K_1,K_2}^{(\ell)} \\
  \end{pmatrix}
  \begin{pmatrix}
    B_1(x_2) \\ \vdots \\ B_{K_2}(x_2)
  \end{pmatrix},
\]

\noindent or simply
$f_\ell = \mathbf{b}_1^t \boldsymbol\Theta_\ell \mathbf{b}_2,$ where
$\mathbf{b}_1$ is a $K_1 \times 1$ vector of B-spline basis functions
evaluated at $x_1$, similarly $\mathbf{b}_2$ is a $K_2 \times 1$
vector of B-spline basis evaluated at $x_2$, and
$\boldsymbol\Theta_\ell$ is a $K_1 \times K_2$ matrix of spline
coefficients for the tensor product approximation of the $\ell$-th
function $f_\ell$. Similarly,
\[
  \frac{\partial}{\partial x_1} f_\ell(x_1, x_2) =
  \begin{pmatrix}
    B_1^\prime(x_1) & \cdots & B_{K_1}^\prime(x_1)
  \end{pmatrix}
  \begin{pmatrix}
    \theta_{1,1}^{(\ell)} & \cdots & \theta_{1,K_2}^{(\ell)} \\
    \vdots & \ddots & \vdots \\
    \theta_{K_1,1}^{(\ell)} & \cdots & \theta_{K_1,K_2}^{(\ell)} \\
  \end{pmatrix}
  \begin{pmatrix}
    B_1(x_2) \\ \vdots \\ B_{K_2}(x_2)
  \end{pmatrix},
\]

\noindent is the partial derivative of $f_\ell$ with respect to $x_1,$
and if we write
$\mathbf{b}_k^\prime = \begin{pmatrix} B_1^\prime(x_k) & \cdots &
  B_{K_1}^\prime(x_k) \end{pmatrix} \!,$ $k = 1,2,$ then
$\partial f_\ell / \partial x_1= (\mathbf{b}^\prime_1)^t
\boldsymbol\Theta_\ell \mathbf{b}_2.$ This allows us to see that
\[
  \begin{aligned}
    |\mathbf{J}| & = ((\mathbf{b}^\prime_1)^t \boldsymbol\Theta_1 \mathbf{b}_2) \cdot (\mathbf{b}_1^t \boldsymbol\Theta_2 \mathbf{b}^\prime_2) - (\mathbf{b}_1^t \boldsymbol\Theta_1 \mathbf{b}^\prime_2) \cdot ((\mathbf{b}^\prime_1)^t \boldsymbol\Theta_2 \mathbf{b}_2) \\
    & = \left [ \left(\mathbf{b}_2 \otimes \mathbf{b}_1^\prime \right)^t \text{vec}(\boldsymbol\Theta_1) \right ] \left [ \left(\mathbf{b}_2^\prime \otimes \mathbf{b}_1 \right)^t \text{vec}(\boldsymbol\Theta_2) \right] - \\
    & {} \qquad \left [ \left(\mathbf{b}_2^\prime \otimes \mathbf{b}_1 \right)^t \text{vec}(\boldsymbol\Theta_1) \right ] \left [ \left(\mathbf{b}_2 \otimes \mathbf{b}_1^\prime \right)^t \text{vec}(\boldsymbol\Theta_2) \right] \\
    & = \text{vec}(\boldsymbol\Theta_1)^t \left [ \left(\mathbf{b}_2 \otimes \mathbf{b}_1^\prime \right) \left(\mathbf{b}_2^\prime \otimes \mathbf{b}_1 \right)^t - \left(\mathbf{b}_2^\prime \otimes \mathbf{b}_1 \right) \left(\mathbf{b}_2 \otimes \mathbf{b}_1^\prime \right)^t \right] \text{vec}(\boldsymbol\Theta_2) \\
    & = \text{vec}(\boldsymbol\Theta_1)^t \mathbf{A}(x_1, x_2)
    \text{vec}(\boldsymbol\Theta_2)
  \end{aligned}
\]

\noindent so the determinant of the Jacobian $\mathbf{J}$ as a
function of $x_1, x_2$ is an inner product of
$\text{vec}(\boldsymbol\Theta_1),$ $\text{vec}(\boldsymbol\Theta_2)$,
weighted by the skew-Symmetric matrix $\mathbf{A}(x_1,x_2).$ Ensuring
the inequality $|\mathbf{J}| > 0$ for all values of $x_1, x_2$ remains
a difficult task, but we can chose a set of basis for which $\mathbf{A}(x_1,x_2)$ does not depend on $x_1,x_2.$

\section{Constrained spatial deformation estimation}\label{sec:cdefest}

We will use an approach similar to \citet{musse2001topology}. Consider B-splines of degree 1 on $[0,T]$. Assume that there are $K-2$
equally spaced inner knots, where
$0 < \tau_1 < \ldots < \tau_{K-2} < T$. Since the knots are equally
spaced, they can be written as
$0 < \tau < 2\tau < \ldots < (K-2)\tau < T,$ where $\tau = T/(K-1).$
In this case, the $K$ B-spline bases are given by
\[
  \begin{aligned}
    B_1(x) & = \begin{cases}1-\dfrac{x}{\tau} & \text{ if } x \in [0, \tau], \\ 0& \text{ otherwise, } \end{cases} \\
    B_2(x) & = \begin{cases}\dfrac{x}{\tau} & \text{ if } x \in [0, \tau], \\ 2 - \dfrac{x}{\tau} & \text{ if } x \in [\tau, 2\tau], \\  0& \text{ otherwise, } \end{cases} \\
    B_k(x) & = B_2(x - (k-2) \tau), \quad k = 3, \ldots, K-1. \\
    B_K(x) & = B_1(K - 1 - x), \\
  \end{aligned}\] with derivatives
\[
  \begin{aligned}
    B^\prime_1(x) & = \begin{cases}-\dfrac{1}{\tau} & \text{ if } x \in [0, \tau], \\ 0& \text{ otherwise, } \end{cases} \\
    B^\prime_2(x) & = \begin{cases}\dfrac{1}{\tau} & \text{ if } x \in
      [0, \tau], \\ - \dfrac{1}{\tau} & \text{ if } x \in [\tau,
      2\tau], \\ 0& \text{ otherwise, } \end{cases}
  \end{aligned}\]

Consider
$(x_1, x_2) \in [\tau_{i-1}, \tau_{i}] \times [\tau_{j-1}, \tau_{j}],$
where $i$ and $j$ is between $1$ and $K-1$ (where $\tau_0 = 0$ and
$\tau_{K-1} = T$). Then there are only 4 bases that evaluate to
non-zero values, indexed by $i-1$, $i$, $j-1$ and $j$, so
\[
  \begin{aligned}
    \mathbf{b}_2 \otimes \mathbf{b}_1^\prime & = \begin{pmatrix} \mathbf{0}^t & \dfrac{x_2-(j-1)\tau}{\tau^2} & -\dfrac{x_2-(j-1)\tau}{\tau^2} & \mathbf{0}^t & -\dfrac{x_2- (j-2)\tau}{\tau^2} & \dfrac{x_2-(j-2)\tau}{\tau^2} & \mathbf{0}^t \end{pmatrix}^t \\
    \mathbf{b}^\prime_2 \otimes \mathbf{b}_1 & = \begin{pmatrix} \mathbf{0}^t & \dfrac{x_1-(i-1)\tau}{\tau^2} & -\dfrac{x_1-(i-2)\tau}{\tau^2} & \mathbf{0}^t  & -\dfrac{x_1-(i-1)\tau}{\tau^2} & \dfrac{x_1-(i-2)\tau}{\tau^2}  & \mathbf{0}^t \end{pmatrix}^t \\
  \end{aligned}
\]
and therefore, looking only at the non-zero pairs
$(i-1, j-1), (i, j-1), (i-1, j), (i,j)$, we have
\[
  \mathbf{A}_{(i-1) : i, (j-1):j}(x_1, x_2) = \frac{1}{\tau^4}
  \begin{pmatrix}
    0 &  a &  b & c \\
    -a &  0 &  d & e \\
    -b & -d &  0 & f \\
    -c & -e & -f & 0 \\
  \end{pmatrix}
\]
where
\[
  \begin{aligned}
    a & = -\tau(x_2 - (j-1) \tau) \\
    b & = \tau(x_1 - (i-1) \tau) \\
    c & = \tau(x_2 - x_1 - \tau(j-i))  \\
    d & = -\tau(x_1 + x_2 - \tau (i + j - 3)) \\
    e & = \tau(x_1 - (i-2) \tau) \\
    f & = -\tau(x_2 - (j-2) \tau)\\
  \end{aligned}
\]
note
$\text{vec}(\boldsymbol\Theta_1)^t\mathbf{A}(x_1, x_2)
\text{vec}(\boldsymbol\Theta_2)$ is therefore proportional to
\[
  \begin{aligned}
    & =  (x_1-(i-1)\tau)\left(\theta_{i-1,j}^{(1)}(\theta_{i,j-1}^{(2)} - \theta_{i-1,j-1}^{(2)}) - \theta_{i-1,j-1}^{(1)}(\theta_{i,j}^{(2)} - \theta_{i-1,j}^{(2)})\right) \\
    & {} \quad + (x_1-(i-2)\tau) \left(\theta_{i,j-1}^{(1)}(\theta_{i,j}^{(2)} - \theta_{i-1,j}^{(2)}) - \theta_{i,j}^{(1)}(\theta_{i,j-1}^{(2)} - \theta_{i-1,j-1}^{(2)})\right)\\
    & {} \quad + (x_2-(j-1)\tau) \left(\theta_{i-1,j-1}^{(1)}(\theta_{i,j}^{(2)} - \theta_{i,j-1}^{(2)}) - \theta_{i,j-1}^{(1)}(\theta_{i-1,j}^{(2)} - \theta_{i-1,j-1}^{(2)})\right)\\
    & {} \quad + (x_2-(j-2)\tau) \left(\theta_{i,j}^{(1)}(\theta_{i-1,j}^{(2)} - \theta_{i-1,j-1}^{(2)}) - \theta_{i-1,j}^{(1)}(\theta_{i,j}^{(2)} - \theta_{i,j-1}^{(2)})\right)\\
  \end{aligned}
\]

The above equations describe a plane in $x_1, x_2$ with coefficients
depending on $\boldsymbol\Theta_1, \boldsymbol\Theta_2$. Now, since
$x_1 \in [\tau_{i-1},\tau_{i}]$ and $x_2 \in [\tau_{j-1},\tau_{j}]$,
where $\tau_i = i \tau,$ $i = 1, 2, \ldots, K-1$ (and similarly for
$j$), we have four restrictions to consider:

\begin{itemize}

\item When $x_1 = (i-1)\tau$ and $x_2 = (j-1)\tau$, \[
    \begin{aligned}
      |\mathbf{J}| & = \tau \left(\theta_{i,j-1}^{(1)}(\theta_{i,j}^{(2)} - \theta_{i-1,j}^{(2)}) - \theta_{i,j}^{(1)}(\theta_{i,j-1}^{(2)} - \theta_{i-1,j-1}^{(2)})\right. \\
      & {} \quad + \left . \theta_{i,j}^{(1)}(\theta_{i-1,j}^{(2)} - \theta_{i-1,j-1}^{(2)}) - \theta_{i-1,j}^{(1)}(\theta_{i,j}^{(2)} - \theta_{i,j-1}^{(2)})\right)\\
    \end{aligned}
  \]

\item When $x_1 = i\tau$ and $x_2 = (j-1)\tau$, \[
    \begin{aligned}
      |\mathbf{J}| & = \tau\left(\theta_{i-1,j}^{(1)}(\theta_{i,j-1}^{(2)} - \theta_{i-1,j-1}^{(2)}) - \theta_{i-1,j-1}^{(1)}(\theta_{i,j}^{(2)} - \theta_{i-1,j}^{(2)})\right) \\
      & {} \quad + 2\tau \left(\theta_{i,j-1}^{(1)}(\theta_{i,j}^{(2)} - \theta_{i-1,j}^{(2)}) - \theta_{i,j}^{(1)}(\theta_{i,j-1}^{(2)} - \theta_{i-1,j-1}^{(2)})\right)\\
      & {} \quad + \tau \left(\theta_{i,j}^{(1)}(\theta_{i-1,j}^{(2)} - \theta_{i-1,j-1}^{(2)}) - \theta_{i-1,j}^{(1)}(\theta_{i,j}^{(2)} - \theta_{i,j-1}^{(2)})\right)\\
    \end{aligned}
  \]

\item When $x_1 = (i-1)\tau$ and $x_2 = j\tau$, \[
    \begin{aligned}
      |\mathbf{J}| & = \tau \left(\theta_{i,j-1}^{(1)}(\theta_{i,j}^{(2)} - \theta_{i-1,j}^{(2)}) - \theta_{i,j}^{(1)}(\theta_{i,j-1}^{(2)} - \theta_{i-1,j-1}^{(2)})\right)\\
      & {} \quad + \tau \left(\theta_{i-1,j-1}^{(1)}(\theta_{i,j}^{(2)} - \theta_{i,j-1}^{(2)}) - \theta_{i,j-1}^{(1)}(\theta_{i-1,j}^{(2)} - \theta_{i-1,j-1}^{(2)})\right)\\
      & {} \quad + 2\tau \left(\theta_{i,j}^{(1)}(\theta_{i-1,j}^{(2)} - \theta_{i-1,j-1}^{(2)}) - \theta_{i-1,j}^{(1)}(\theta_{i,j}^{(2)} - \theta_{i,j-1}^{(2)})\right)\\
    \end{aligned}
  \]

\item When $x_1 = i\tau$ and $x_2 = j\tau$, \[
    \begin{aligned}
      |\mathbf{J}| & = \tau\left(\theta_{i-1,j}^{(1)}(\theta_{i,j-1}^{(2)} - \theta_{i-1,j-1}^{(2)}) - \theta_{i-1,j-1}^{(1)}(\theta_{i,j}^{(2)} - \theta_{i-1,j}^{(2)})\right) \\
      & {} \quad + 2\tau \left(\theta_{i,j-1}^{(1)}(\theta_{i,j}^{(2)} - \theta_{i-1,j}^{(2)}) - \theta_{i,j}^{(1)}(\theta_{i,j-1}^{(2)} - \theta_{i-1,j-1}^{(2)})\right)\\
      & {} \quad + \tau \left(\theta_{i-1,j-1}^{(1)}(\theta_{i,j}^{(2)} - \theta_{i,j-1}^{(2)}) - \theta_{i,j-1}^{(1)}(\theta_{i-1,j}^{(2)} - \theta_{i-1,j-1}^{(2)})\right)\\
      & {} \quad + 2\tau \left(\theta_{i,j}^{(1)}(\theta_{i-1,j}^{(2)} - \theta_{i-1,j-1}^{(2)}) - \theta_{i-1,j}^{(1)}(\theta_{i,j}^{(2)} - \theta_{i,j-1}^{(2)})\right)\\
    \end{aligned}
  \]

\end{itemize}


This collection of constraints, for $i=1,\ldots,K-1$ and $j=1, \ldots, K-1$ together imply in a non-folding deformation map, and can be enforced with constrained optimization routines. We have employed \citet{svanberg2002class} constrained optimization algorithm, avaliable in the \texttt{nloptr} package \citep{johnson2020nloptr}. The code is available as an \texttt{R} package in \url{https://github.com/guiludwig/bsplinedef}.

\section{Simulation study}\label{sec:sim}

\begin{figure}[!tb]
    \centering
    \includegraphics[width = .9\textwidth]{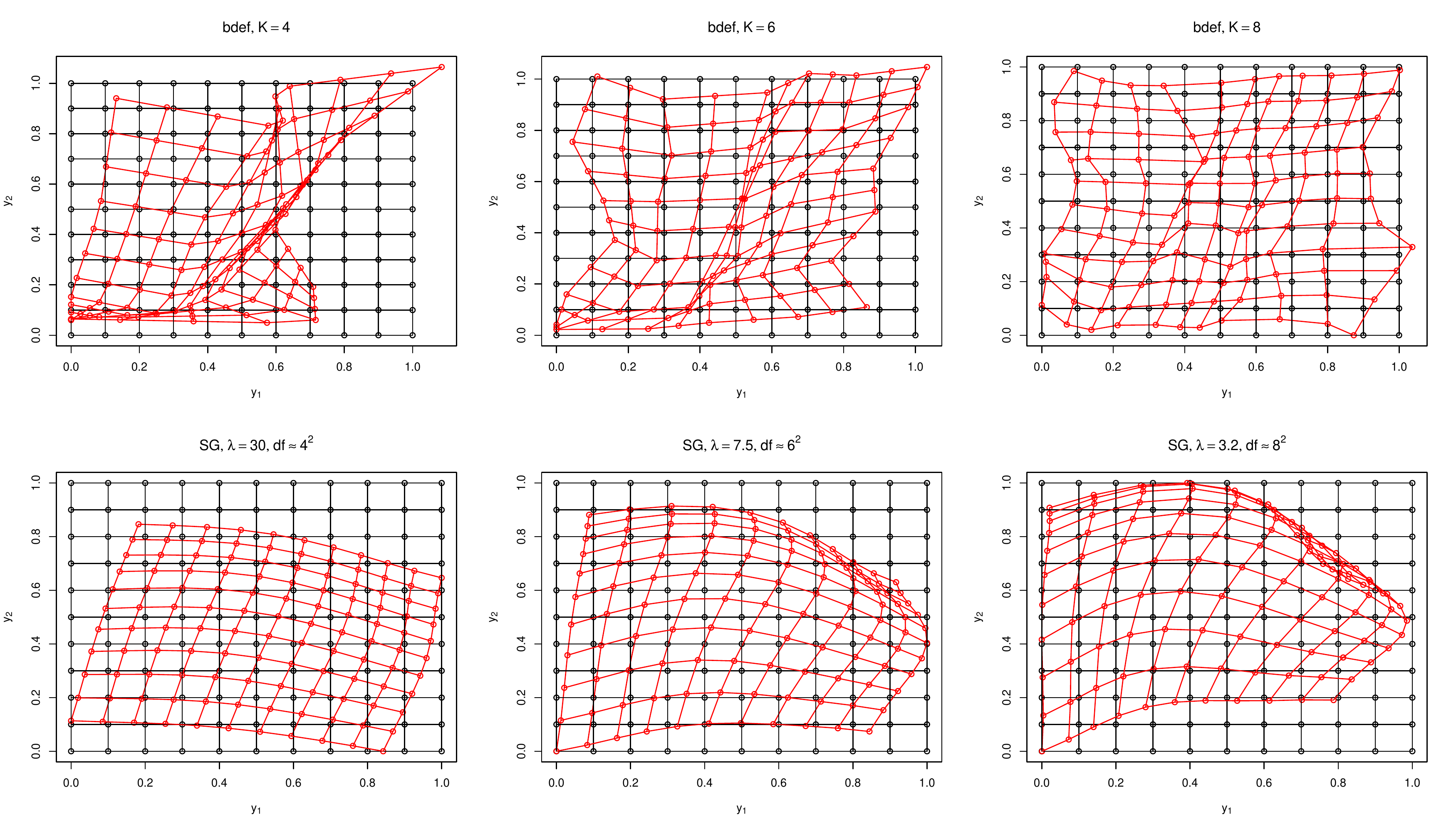}
    \caption{Comparison of estimated deformation functions for simulated data. The upper row corresponds to the proposed regularized B-spline approach with $K=4,6$ and $8,$ respectively. The bottom row corresponds to the \citet{sampson1992nonparametric} method with $\lambda = 30,$ $7.5$ and $3.2,$ which are equivalent in degrees of freedom to the upper cases.}
    \label{fig:sim1}
\end{figure}

To evaluate the performance of the algorithm, we conducted a simulation study based on the swirl function shown in Figure \ref{fig:example1}. Each sample has $n=121$ spatial points on a regular grid in the geographical domain $\mathcal{G} = [0,1]\times[0,1].$ We simulated from a Gaussian random field with mean function $\mu(\mathbf{x}) = 0$ and covariance function $C(\mathbf{x}_1, \mathbf{x}_2) = \tau^2 + \sigma^2 e^{-\|\mathbf{y}_1 - \mathbf{y}_2\|/\phi},$ with parameters $\tau^2 = \sigma^2 = 1,$ and $\phi = 0.25.$ First, consider a single realization of the spatial random field. The estimated deformation maps are shown in Figure \ref{fig:sim1}. We have obtained the constrained B-spline deformations (hereafter, \texttt{bdef}) with $K \times K =4^2,6^2$ and $8^2$ basis functions. The estimated deformation function does not fold, even though the true deformation function is difficult to be recovered. The case when $K=8$ performs better than $K=4$ or $K=6,$ indicating that there are features in the deformation map that need  a large number of degrees of freedom to be estimated. On the other hand, the functions using \citet{sampson1992nonparametric} (\texttt{SG}) method have smoothing parameters $\lambda = 30,$ $7.5$ and $3.2,$  set to match the \texttt{bdef} approach. They start showing folding at $\lambda = 3.2,$ and cannot recover deformation maps that require a number of degrees of freedom larger than $8^2.$ Note that the \texttt{SG} maps were stretched or shrunk to fit the plot area, but the \texttt{bdef} maps did not require this step.

\begin{figure}[!tb]
    \centering
    \includegraphics[width = .9\textwidth]{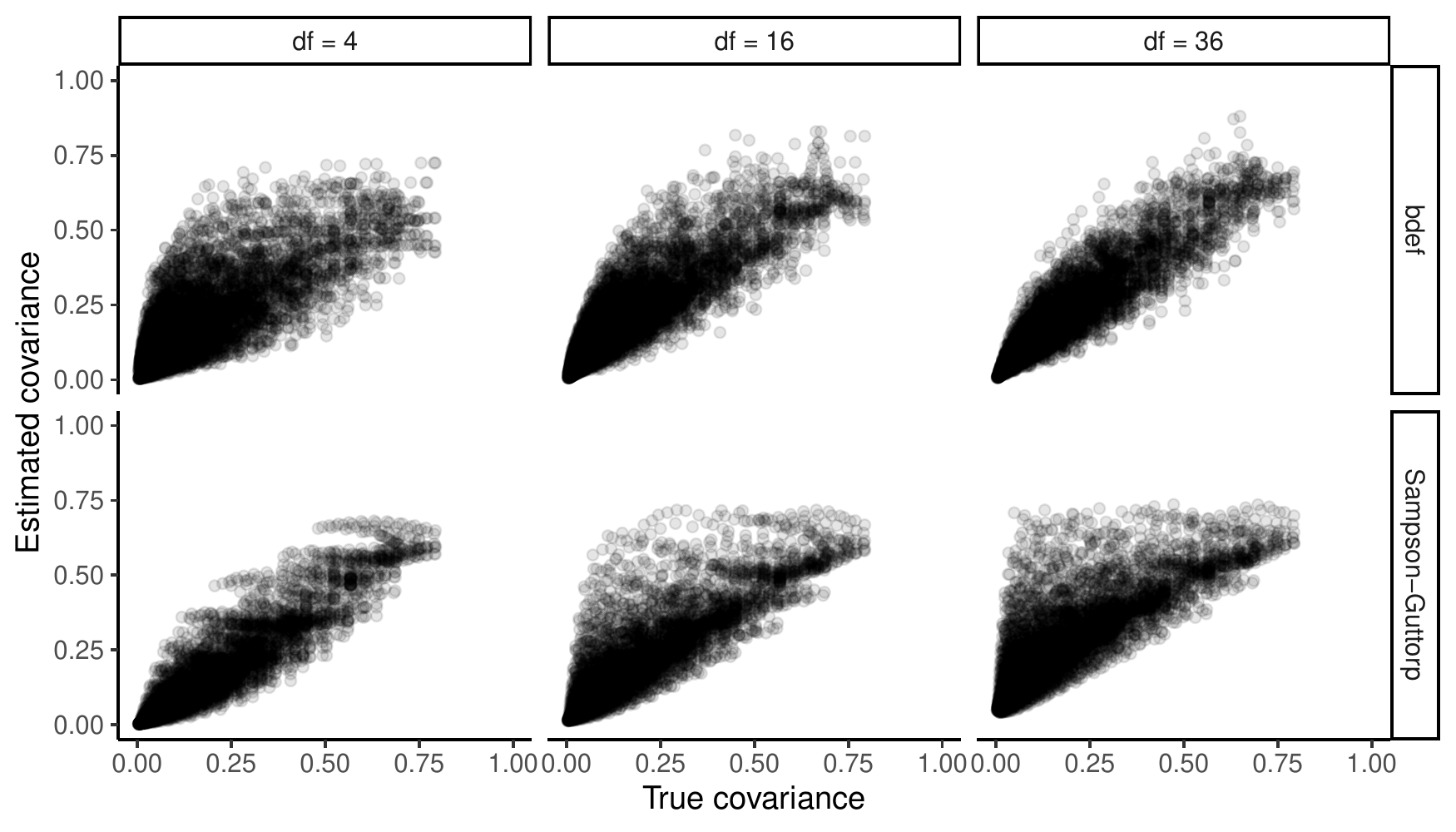}
    \caption{Comparison of upper-diagonal entries for the estimated covariance matrix of the simulated data, versus the true covariance matrix. The upper row corresponds to the proposed regularized B-spline approach with $K=4,6$ and $8,$ respectively. The bottom row corresponds to the \citet{sampson1992nonparametric} method with $\lambda = 30,$ $7.5$ and $3.2,$ which are equivalent in degrees of freedom to the upper cases.}
    \label{fig:sim2}
\end{figure}

The comparison of estimated covariance matrices allow us to overlook the identifiability issues with rotations, shifts and scaling of the estimated maps. In Figure \ref{fig:sim2} we show a scatterplot of the upper-diagonal entries of the estimated covariance matrices for the data, versus the true covariance matrices. We remark that the \texttt{bdef} method shows no apparent bias and becomes more accurate as the number of degrees of freedom increase. On the other hand, \texttt{SG} has good performance at a small number of degrees of freedom (more smoothing), but tends to overestimate matrix entries as the number of degrees of freedom for the thin-plate spline increases.

\section{Case study: Rainfall data in southeastern Brazil}\label{sec:app}

The dataset we use to illustrate our method comes from  meteorological surveys conducted by INMET -- Instituto Nacional de   Meteorologia, Brazil. The measurements are made at every 15 minutes, and daily accumulated values are made available. Following \citet{rozante2010combining}, we grouped the rainfall data in periods of 10 days.

Since we seek temporally stationary data, we decided to restrain the data collection to 2018-01-01 to 2018-03-30 (rainfall season), for a total of 9 time periods (period 1: 2018-01-01 to 2018-01-10, ..., period 9: 2018-03-22 to 2018-03-31).
    
\begin{figure}
    \centering
    \includegraphics[scale = .45]{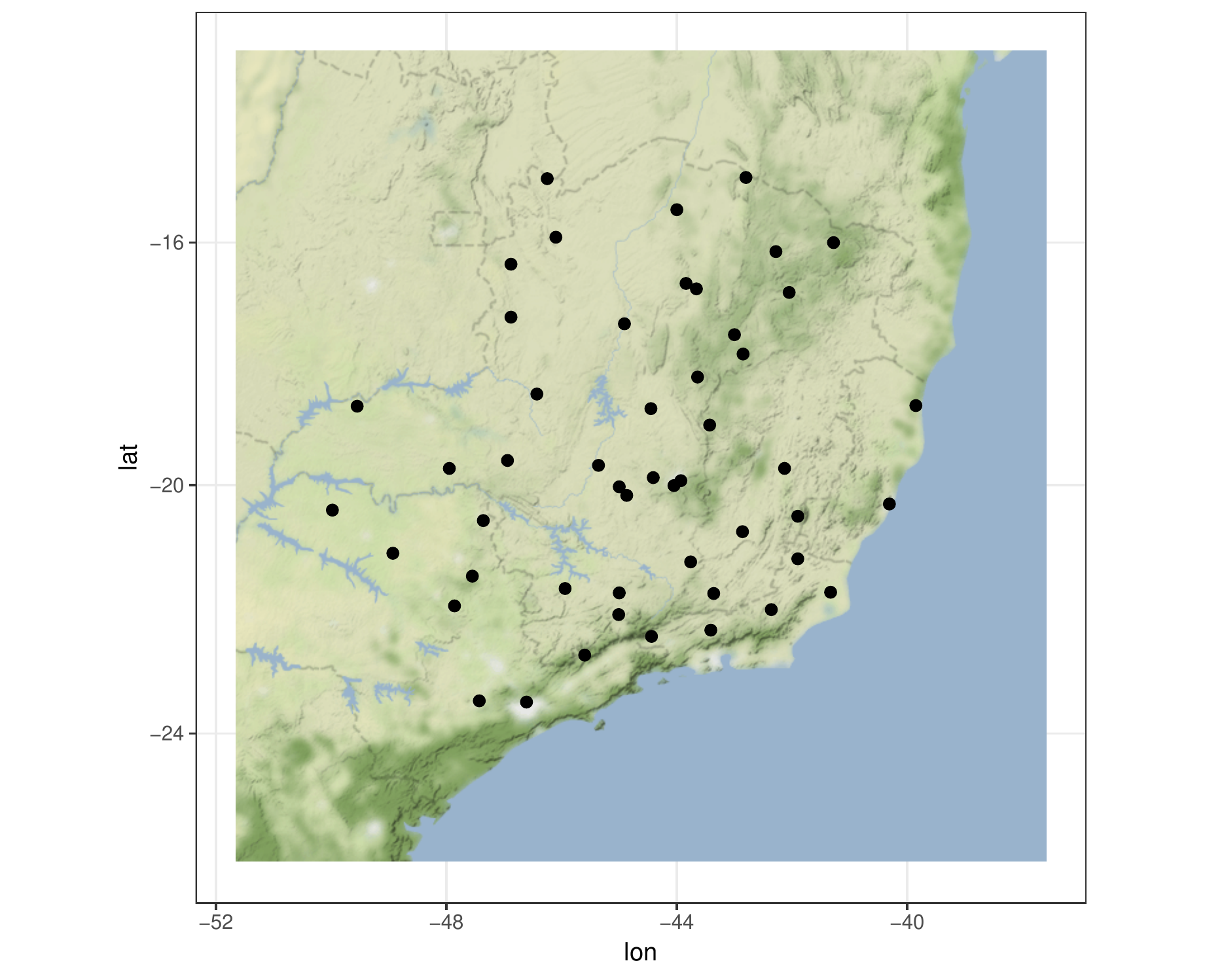}
    \includegraphics[scale = .4]{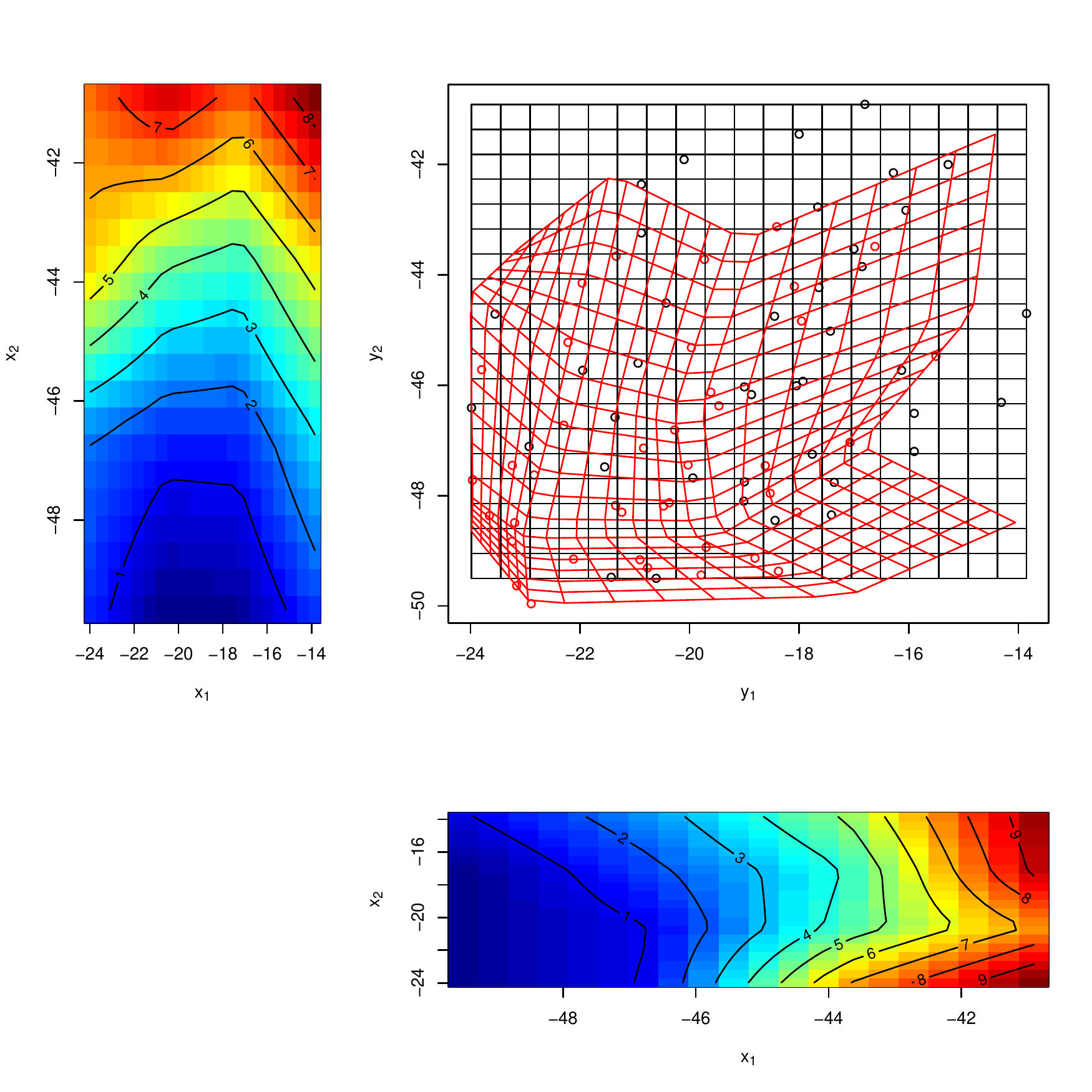}
    \caption{Left panel: map of meteorological stations located in southeastern Brazil. Right panel: estimated deformation functions using $K\times K = 16$ B-spline basis.}
    \label{fig:realDataLocations}
\end{figure}

We selected the 50 stations of the southeastern region of Brazil that had complete observations available during the period. The stations are shown in the left panel of Figure \ref{fig:realDataLocations}. Data can be obtained at \url{http://www.inmet.gov.br/portal/index.php?r=bdmep/bdmep}.

In the right panel of Figure \ref{fig:realDataLocations}, we show the estimated deformation map, using $K \times K = 16$ B-spline basis functions. The estimated deformation map reveals topographical features shown in the left panel of the same Figure. Weather stations located in northwestern flat lands of Minas Gerais are treated as closer to each other than stations near the rough coast of Rio de Janeiro and S\~ao Paulo states, where elevation changes are abrupt.


\begin{figure}
    \centering
    \includegraphics[scale = .4]{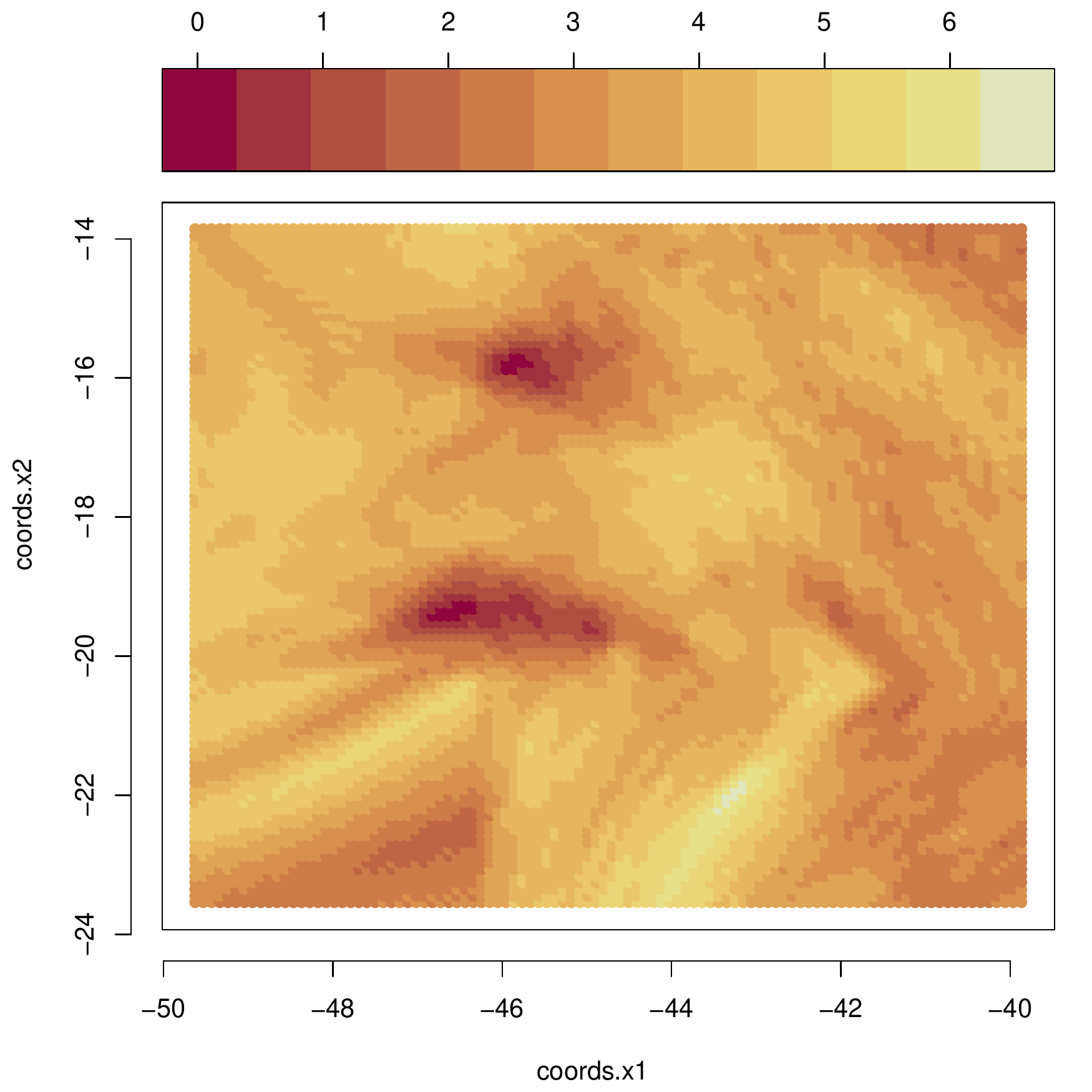} 
    \includegraphics[scale = .4]{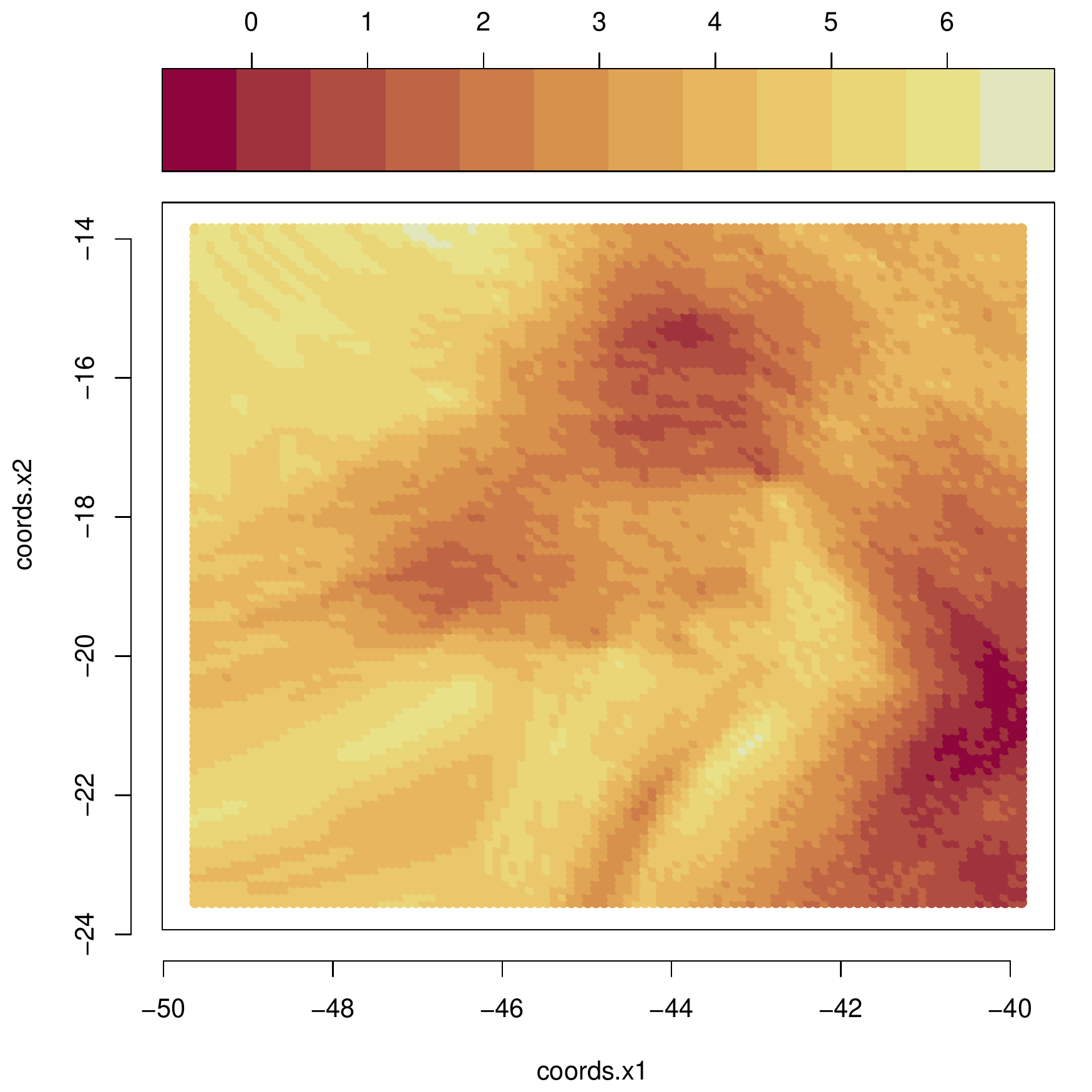}    
    \caption{Conditional simulation of the Gaussian random field (Kriging) for the 10-day periods starting in 2018--01--01 and 2018--01--11.}
    \label{fig:realDataSim}
\end{figure}

In Figure \ref{fig:realDataSim} we perform conditional simulation of the Gaussian random field (Kriging) for the 10-day periods starting in 2018--01--01 and 2018--01--11, showing the resulting prediction maps.


\section*{Acknowledgments}

Ronaldo Dias was supported by the FAPESP foundation, grant 2018/04654-9. Guilherme Ludwig was supported by FAPESP grants 2018/04654-9 and 2019/03517-0, as well as FAEPEX grant 2253/19.

\bibliographystyle{abbrvnat}
\bibliography{references}

\appendix

\section{Derivation of the non-folding constraint}

We can derive the constraint presented in Section \ref{sec:cdefest} as follows: Observe that

 \[
   \begin{aligned}
    \mathbf{b}_2 \otimes \mathbf{b}_1^\prime & = \begin{pmatrix}
      \mathbf{0}^t & B_{j-1} (x_2) B_{i-1}^\prime (x_1) & B_{j-1}
      (x_2) B_{i}^\prime (x_1) & \mathbf{0}^t & B_{j} (x_2)
      B_{i-1}^\prime (x_1) & B_{j} (x_2) B_{i}^\prime (x_1) &
      \mathbf{0}^t \end{pmatrix}^t \\
    \mathbf{b}^\prime_2 \otimes \mathbf{b}_1 & = \begin{pmatrix} \mathbf{0}^t & B^\prime_{j-1} (x_2)  B_{i-1} (x_1) & B^\prime_{j-1} (x_2)  B_{i} (x_1) & \mathbf{0}^t & B^\prime_{j} (x_2)  B_{i-1} (x_1) & B^\prime_{j} (x_2)  B_{i} (x_1) & \mathbf{0}^t \end{pmatrix}^t \\
  \end{aligned}
\]
 or
 \[
 \begin{aligned}
    \mathbf{b}_2 \otimes \mathbf{b}_1^\prime & = 
    \left(
        \begin{matrix} \mathbf{0}^t & -\left(2 - \dfrac{x_2-(j-3)\tau}{\tau}\right) \dfrac{1}{\tau} & \left(2 - \dfrac{x_2-(j-3)\tau}{\tau}\right) \dfrac{1}{\tau}  \end{matrix} \right. \\
      & \qquad  \left. \begin{matrix} \mathbf{0}^t  & -\dfrac{x_2- (j-2)\tau}{\tau}\dfrac{1}{\tau} & \dfrac{x_2-(j-2)\tau}{\tau}\dfrac{1}{\tau} & \mathbf{0}^t \end{matrix} \right)^t \\ 
    \mathbf{b}^\prime_2 \otimes \mathbf{b}_1 & = 
    \left( \begin{matrix} \mathbf{0}^t & -\left(2 - \dfrac{x_1-(i-3)\tau}{\tau}\right) \dfrac{1}{\tau} & - \dfrac{x_1-(i-2)\tau}{\tau} \dfrac{1}{\tau} \end{matrix} \right. \\
      & \qquad \left. \begin{matrix} \mathbf{0}^t  & \left(2 - \dfrac{x_1-(i-3)\tau}{\tau}\right) \dfrac{1}{\tau} & \dfrac{x_1-(i-2)\tau}{\tau} \dfrac{1}{\tau} & \mathbf{0}^t \end{matrix} \right)^t \\ 
  \end{aligned}
\]
Further simplifying leads to
\[
  \begin{aligned}
    \mathbf{b}_2 \otimes \mathbf{b}_1^\prime & = \begin{pmatrix} \mathbf{0}^t & \dfrac{x_2-(j-1)\tau}{\tau^2} & -\dfrac{x_2-(j-1)\tau}{\tau^2} & \mathbf{0}^t & -\dfrac{x_2- (j-2)\tau}{\tau^2} & \dfrac{x_2-(j-2)\tau}{\tau^2} & \mathbf{0}^t \end{pmatrix}^t \\
    \mathbf{b}^\prime_2 \otimes \mathbf{b}_1 & = \begin{pmatrix} \mathbf{0}^t & \dfrac{x_1-(i-1)\tau}{\tau^2} & -\dfrac{x_1-(i-2)\tau}{\tau^2} & \mathbf{0}^t  & -\dfrac{x_1-(i-1)\tau}{\tau^2} & \dfrac{x_1-(i-2)\tau}{\tau^2}  & \mathbf{0}^t \end{pmatrix}^t \\
  \end{aligned}
\]
and therefore, looking only at the non-zero pairs
$(i-1, j-1), (i, j-1), (i-1, j), (i,j)$, we have
\[
  \mathbf{A}_{(i-1) : i, (j-1):j}(x_1, x_2) = \frac{1}{\tau^4}
  \begin{pmatrix}
    0 &  a &  b & c \\
    -a &  0 &  d & e \\
    -b & -d &  0 & f \\
    -c & -e & -f & 0 \\
  \end{pmatrix}
\]
where
\[
  \begin{aligned}
    a & = -\tau(x_2 - (j-1) \tau) \\
    b & = \tau(x_1 - (i-1) \tau) \\
    c & = \tau(x_2 - x_1 - \tau(j-i))  \\
    d & = -\tau(x_1 + x_2 - \tau (i + j - 3)) \\
    e & = \tau(x_1 - (i-2) \tau) \\
    f & = -\tau(x_2 - (j-2) \tau)\\
  \end{aligned}
\]
note
$\text{vec}(\boldsymbol\Theta_1)^t\mathbf{A}(x_1, x_2)
\text{vec}(\boldsymbol\Theta_2)$ is therefore proportional to
\[
  \frac{1}{\tau}
  \begin{pmatrix}
    \theta_{i-1,j-1}^{(1)} & \theta_{i,j-1}^{(1)} &
    \theta_{i-1,j}^{(1)} & \theta_{i,j}^{(1)}
  \end{pmatrix} \begin{pmatrix}
    \mathbf{0} &                 a &                  b & c\\
    -a & \mathbf{0} &                  d & e\\
    -b &                -d & \mathbf{0}  & f \\
    -c &               -e &                 -f & \mathbf{0} \\
  \end{pmatrix} \begin{pmatrix}
    \theta_{i-1,j-1}^{(2)} \\
    \theta_{i,j-1}^{(2)} \\
    \theta_{i-1,j}^{(2)} \\
    \theta_{i,j}^{(2)} \\
  \end{pmatrix} =
\]
\[
  \boldsymbol\theta^{(1),t} \begin{pmatrix}
    -(x_2 - (j-1) \tau)\theta_{i,j-1}^{(2)} + (x_1 - (i-1) \tau)\theta_{i-1,j}^{(2)} + (x_2 - x_1 - \tau(j-i)) \theta_{i,j}^{(2)} \\
    (x_2 - (j-1) \tau)\theta_{i-1,j-1}^{(2)} -(x_1 + x_2 - \tau (i + j - 3)) \theta_{i-1,j}^{(2)} + (x_1 - (i-2) \tau) \theta_{i,j}^{(2)}\\
    -(x_1 - (i-1) \tau) \theta_{i-1,j-1}^{(2)} + (x_1 + x_2 - \tau (i + j - 3)) \theta_{i,j-1}^{(2)} -(x_2 - (j-2) \tau) \theta_{i,j}^{(2)} \\
    -(x_2 - x_1 - \tau(j-i))\theta_{i-1,j-1}^{(2)} - (x_1 - (i-2) \tau)\theta_{i,j-1}^{(2)} + (x_2 - (j-2) \tau) \theta_{i-1,j}^{(2)} \\
  \end{pmatrix}
\]
\noindent where $\boldsymbol\theta^{(1),t}=\text{vec}(\boldsymbol\Theta_1)^t_{(i-1): i, (j-1):j}$, and


\[
  \boldsymbol\theta^{(1),t} \begin{pmatrix}
    -(x_1-(i-1)\tau)(\theta_{i,j}^{(2)} - \theta_{i-1,j}^{(2)}) + (x_2 - (j-1)\tau)(\theta_{i,j}^{(2)} - \theta_{i,j-1}^{(2)}) \\
    (x_1-(i-2)\tau)(\theta_{i,j}^{(2)}-\theta_{i-1,j}^{(2)}) - (x_2 - (j-1)\tau)(\theta_{i-1,j}^{(2)}-\theta_{i-1,j-1}^{(2)}) \\
    (x_1-(i-1)\tau)(\theta_{i,j-1}^{(2)} - \theta_{i-1,j-1}^{(2)}) - (x_2-(j-2)\tau)(\theta_{i,j}^{(2)} - \theta_{i,j-1}^{(2)}) \\
    -(x_1-(i-2)\tau)(\theta_{i,j-1}^{(2)} - \theta_{i-1,j-1}^{(2)}) + (x_2-(j-2)\tau)(\theta_{i-1,j}^{(2)} - \theta_{i-1,j-1}^{(2)}) \\
  \end{pmatrix} =
\]

\[
  \begin{aligned}
    & =  (x_1-(i-1)\tau)\left(\theta_{i-1,j}^{(1)}(\theta_{i,j-1}^{(2)} - \theta_{i-1,j-1}^{(2)}) - \theta_{i-1,j-1}^{(1)}(\theta_{i,j}^{(2)} - \theta_{i-1,j}^{(2)})\right) \\
    & {} \quad + (x_1-(i-2)\tau) \left(\theta_{i,j-1}^{(1)}(\theta_{i,j}^{(2)} - \theta_{i-1,j}^{(2)}) - \theta_{i,j}^{(1)}(\theta_{i,j-1}^{(2)} - \theta_{i-1,j-1}^{(2)})\right)\\
    & {} \quad + (x_2-(j-1)\tau) \left(\theta_{i-1,j-1}^{(1)}(\theta_{i,j}^{(2)} - \theta_{i,j-1}^{(2)}) - \theta_{i,j-1}^{(1)}(\theta_{i-1,j}^{(2)} - \theta_{i-1,j-1}^{(2)})\right)\\
    & {} \quad + (x_2-(j-2)\tau) \left(\theta_{i,j}^{(1)}(\theta_{i-1,j}^{(2)} - \theta_{i-1,j-1}^{(2)}) - \theta_{i-1,j}^{(1)}(\theta_{i,j}^{(2)} - \theta_{i,j-1}^{(2)})\right)\\
  \end{aligned}
\]

The above equations describe a plane in $x_1, x_2$ with coefficients
depending on $\boldsymbol\Theta_1, \boldsymbol\Theta_2$. Now, since
$x_1 \in [\tau_{i-1},\tau_{i}]$ and $x_2 \in [\tau_{j-1},\tau_{j}]$,
where $\tau_i = i \tau,$ $i = 1, 2, \ldots, K-1$ (and similarly for
$j$), we have four restrictions to consider:

\begin{itemize}

\item When $x_1 = (i-1)\tau$ and $x_2 = (j-1)\tau$, \[
    \begin{aligned}
      |\mathbf{J}| & = \tau \left(\theta_{i,j-1}^{(1)}(\theta_{i,j}^{(2)} - \theta_{i-1,j}^{(2)}) - \theta_{i,j}^{(1)}(\theta_{i,j-1}^{(2)} - \theta_{i-1,j-1}^{(2)})\right. \\
      & {} \quad + \left . \theta_{i,j}^{(1)}(\theta_{i-1,j}^{(2)} - \theta_{i-1,j-1}^{(2)}) - \theta_{i-1,j}^{(1)}(\theta_{i,j}^{(2)} - \theta_{i,j-1}^{(2)})\right)\\
    \end{aligned}
  \]

\item When $x_1 = i\tau$ and $x_2 = (j-1)\tau$, \[
    \begin{aligned}
      |\mathbf{J}| & = \tau\left(\theta_{i-1,j}^{(1)}(\theta_{i,j-1}^{(2)} - \theta_{i-1,j-1}^{(2)}) - \theta_{i-1,j-1}^{(1)}(\theta_{i,j}^{(2)} - \theta_{i-1,j}^{(2)})\right) \\
      & {} \quad + 2\tau \left(\theta_{i,j-1}^{(1)}(\theta_{i,j}^{(2)} - \theta_{i-1,j}^{(2)}) - \theta_{i,j}^{(1)}(\theta_{i,j-1}^{(2)} - \theta_{i-1,j-1}^{(2)})\right)\\
      & {} \quad + \tau \left(\theta_{i,j}^{(1)}(\theta_{i-1,j}^{(2)} - \theta_{i-1,j-1}^{(2)}) - \theta_{i-1,j}^{(1)}(\theta_{i,j}^{(2)} - \theta_{i,j-1}^{(2)})\right)\\
    \end{aligned}
  \]

\item When $x_1 = (i-1)\tau$ and $x_2 = j\tau$, \[
    \begin{aligned}
      |\mathbf{J}| & = \tau \left(\theta_{i,j-1}^{(1)}(\theta_{i,j}^{(2)} - \theta_{i-1,j}^{(2)}) - \theta_{i,j}^{(1)}(\theta_{i,j-1}^{(2)} - \theta_{i-1,j-1}^{(2)})\right)\\
      & {} \quad + \tau \left(\theta_{i-1,j-1}^{(1)}(\theta_{i,j}^{(2)} - \theta_{i,j-1}^{(2)}) - \theta_{i,j-1}^{(1)}(\theta_{i-1,j}^{(2)} - \theta_{i-1,j-1}^{(2)})\right)\\
      & {} \quad + 2\tau \left(\theta_{i,j}^{(1)}(\theta_{i-1,j}^{(2)} - \theta_{i-1,j-1}^{(2)}) - \theta_{i-1,j}^{(1)}(\theta_{i,j}^{(2)} - \theta_{i,j-1}^{(2)})\right)\\
    \end{aligned}
  \]

\item When $x_1 = i\tau$ and $x_2 = j\tau$, \[
    \begin{aligned}
      |\mathbf{J}| & = \tau\left(\theta_{i-1,j}^{(1)}(\theta_{i,j-1}^{(2)} - \theta_{i-1,j-1}^{(2)}) - \theta_{i-1,j-1}^{(1)}(\theta_{i,j}^{(2)} - \theta_{i-1,j}^{(2)})\right) \\
      & {} \quad + 2\tau \left(\theta_{i,j-1}^{(1)}(\theta_{i,j}^{(2)} - \theta_{i-1,j}^{(2)}) - \theta_{i,j}^{(1)}(\theta_{i,j-1}^{(2)} - \theta_{i-1,j-1}^{(2)})\right)\\
      & {} \quad + \tau \left(\theta_{i-1,j-1}^{(1)}(\theta_{i,j}^{(2)} - \theta_{i,j-1}^{(2)}) - \theta_{i,j-1}^{(1)}(\theta_{i-1,j}^{(2)} - \theta_{i-1,j-1}^{(2)})\right)\\
      & {} \quad + 2\tau \left(\theta_{i,j}^{(1)}(\theta_{i-1,j}^{(2)} - \theta_{i-1,j-1}^{(2)}) - \theta_{i-1,j}^{(1)}(\theta_{i,j}^{(2)} - \theta_{i,j-1}^{(2)})\right)\\
    \end{aligned}
  \]

\end{itemize}

\end{document}